# Direct Investigation of the Birefringent Optical Properties of Black Phosphorus with Picosecond Interferometry

*Wei Zheng, Andrei Nemilentsau, Dustin Lattery, Peipei Wang, Tony Low, Jie Zhu,[*] and Xiaojia Wang[*]*

W. Zheng, D. Lattery, Dr. J. Zhu, Prof. X. Wang
Department of Mechanical Engineering, University of Minnesota, Minneapolis, MN 55455, USA
E-mail: zhuj@umn.edu & wang4940@umn.edu

Dr. A. Nemilentsau, Prof. T. Low
Department of Electrical and Computer Engineering, University of Minnesota, Minneapolis, MN 55455, USA

Dr. P. Wang
Department of Physics, Southern University of Science and Technology, Shenzhen, Guangdong 518055, China

**Abstract:** Black phosphorus (BP) is an emerging two-dimensional semiconducting material with great potential for nanoelectronic and nanophotonic applications, especially owing to its unique anisotropic electrical and optical properties. Many theoretical studies have predicted the anisotropic optical properties of BP, but the direct experimental quantification remains challenging. The difficulties stem from the ease of BP's degradation when exposed to air in ambient conditions, and from the indirect nature of conventional approaches that are subject to large measurement uncertainties. This work reports a direct investigation of the birefringent optical constants of micrometer-thick BP samples with picosecond (ps) interferometry, over the wavelength range from 780 to 890 nm. In this ps-interferometry approach, an ultrathin (~5 nm) platinum layer for launching acoustic waves naturally protects the BP flake from degradation. The birefringent optical constants of BP for light polarization along the two primary crystalline orientations, zigzag and armchair, are directly obtained via fitting the attenuated Brillouin scattering signals. A bi-exponential model is further proposed to analyze the BS signals for a random incident light polarization. The BP experimental results and the associated measurement sensitivity analysis demonstrate the reliability and accuracy of the ps-interferometry approach for capturing the polarization-dependent optical properties of birefringent materials.

TOC

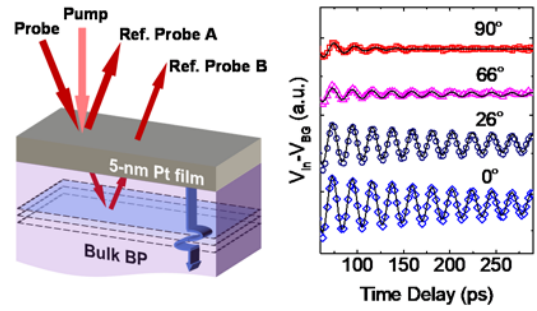

Black phosphorus (BP) is an emerging two-dimensional semiconducting material with unique anisotropic optical properties. This work reports a direct investigation of the birefringent optical constants of BP with picosecond interferometry in the near-infrared regime. The birefringent optical constants of BP for light polarization along zigzag and armchair directions are obtained via analyzing the attenuated Brillouin scattering signals.

# 1. Introduction

Black phosphorus (BP) has become increasingly promising as an emerging two-dimensional (2D) material with great potential for optical and electrical applications.[1-10] It is formed as a stack of parallel 2D sheets with adjacent layers under van der Waals interactions.[11-14] This structure makes BP easy to be mechanically exfoliated into monolayer or few-layer BP flakes, allowing for its integration with nanoscale devices.[1,4,15-16] The bandgap energy of BP depends on its number of layers and can be tuned from 0.3 (bulk BP) to 1.0 eV (monolayer).[17-18] BP has an orthorhombic structure with in-plane covalent bonding, with its characteristic zigzag (ZZ) and armchair (AC) crystal orientations. The unique atomic puckering crystal structure, resulting from the sp$^3$ nonequivalent hybridization of each P atom, leads to the robust in-plane anisotropy nature of BP.[19] Previous studies have shown strong anisotropy of BP in its physical properties, along high-symmetry ZZ and AC directions, including thermal,[20-23] electrical,[24-25] and optical properties.[3,19,26-29] The bandgap tunability, high mobility[4] and anisotropic nature of BP makes it a potential competitor against other well-studied 2D materials, such as graphene and molybdenum disulfide,[30-33] as a promising candidate for next-generation nanoelectronic and nanophotonic devices. In addition, conventional birefringent optics are typically bulky due to the inherently weak birefringence. Black phosphorus, in conjunction with plasmonic effects, would also enable strong tunable birefringence with atomically thin material.[34] Especially, the birefringent mid-infrared optical response in black phosphorus is exceptionally large compared to conventional birefringent non-cubic crystals such as barium borate, calcite, ruby etc., which only features a differential optical index of ~0.1[9,24]. Birefringence is used in many optical devices,[35] such as liquid crystal displays, where the display brightness is controlled via polarized light through a birefringent material. Another example is wave plates which are widely used in free-space optics measurement.

To advance the future potential of BP for nanophotonic devices,[2-5,36-37] precise characterization of the birefringent optical constants of BP crystals is of great importance. To date, there have been literature studies reporting the optical properties of BP, both theoretical[24,38-41] and experimental.[9,19,28-30] These studies have indicated one unique feature of BP, that its extinction coefficient ($\kappa$, the imaginary part of the complex refractive index), is strongly dependent on its crystalline orientation. The absorption is the strongest when the incident light is polarized along the AC direction, and the weakest when the light is polarized along the ZZ direction.[24,38-41]

Experimental investigation of BP is challenging because of its degradable nature when exposed to air. As a result, conventional spectroscopic studies usually require special controlled environments or protective approaches.[27,30] For example, in the spectroscopic and phase-anisotropy measurements, BP thin films were supported on a transparent substrate (glass or sapphire) and coated with a thin layer of Poly(methyl methacrylate) (PMMA) for protection.[27] Both the substrates and protective layers would also contribute to the reflection and transmission signals, and therefore induce more complexities and uncertainties in analyzing the experimental data. Other than that, in spectroscopic measurements, BP samples need to be ultrathin (tens of nanometers) for sufficient light transmission, since BP is highly absorbing for certain light polarization.[3,19,27,30] As an indirect approach for obtaining the optical constants ($\kappa$ and $n$, the real part of the refractive index), conventional spectroscopic methods need precise information about the thickness of BP and the protective/supporting layers. This will introduce not only additional thickness characterizations but also measurement uncertainties in the extracted optical constants.[39] Another indirect method based on the electron energy loss spectroscopy (EELS), has also been employed for optically characterizing 100~200-nm-thick BP films.[29] In this method, a high vacuum chamber at the temperature of ~20 K was applied to minimize thermal broadening in signals, with the assumption that the measured results would not be temperature dependent.[29] These aforementioned challenges in previous studies make it difficult to directly and precisely obtain the birefringent optical constants of BP.

In this study, we synthesize high-quality large crystals of BP and employ picosecond (ps) interferometry to study the birefringent in-plane optical properties of BP that are micrometers in thickness. Our ps-interferometry method is upgraded from a basic ultrafast all-optical pump-probe setup,[42] which captures the Brillouin scattering (BS) signals from the acoustic wave propagation within the micrometer-thick BP. The interference of probe beams, reflected from the BP surface and from the moving acoustic wave front inside of BP, allows for the direct manifestation of the birefringent optical absorption of BP from the attenuated BS signals. We obtain reliable values of the polarization-dependent optical constants of BP by coating each BP flake with a 5-nm platinum (Pt) transducer, which is used to launch acoustic waves and naturally prevents the contamination and degradation of the BP crystal.

## 2. Picosecond Interferometry

The birefringent in-plane optical constants of BP flakes are characterized with ps interferometry, which is a modified pump-probe method for measuring acoustic wave propagation and attenuation.[43-46] Prior to ps-interferometry measurements, each BP sample is coated with a semitransparent thin film. In this work, BP flakes with size on the order of a few microns are mechanically exfoliated and placed on silicon wafers. Figure 1a depicts the layered structure of BP, with the three crystalline directions denoted as *a* (zigzag or ZZ), *b* (interlayer or through-plane), and *c* (armchair or AC). A 5-nm Pt layer is deposited on the surface of BP flake to act as the transducer.

## 2.1. Brillouin Scattering Signals

In our ps-interferometry measurements, when the pump beam reaches the Pt transducer, the laser energy is deposited to free electrons in Pt. The hot electrons pass their energy to the lattice of the Pt transducer, which becomes uniformly distributed throughout the entire transducer at a time scale of a few picoseconds.[47] This leads to a sudden temperature rise of the Pt transducer under the impulse pump heating, which causes the thermal expansion of Pt. The deformation of the Pt transducer under a thermal stress depends linearly on its bulk modulus, its thermal expansion coefficient, and the temperature rise. Subsequently, a coherent strain pulse is formed inside the Pt transducer, which travels downwards into the BP substrate beneath at the speed of sound in BP along the interlayer direction.[45]

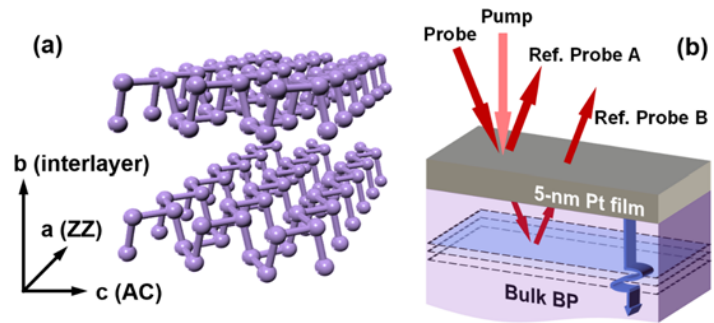

**Figure 1.** (a) A schematic of the BP crystalline structure. The arrows *a*, *b*, and *c* indicate ZZ, interlayer, and AC crystalline directions. (b) Brillouin scattering schematic diagram. The pump beam launches longitudinal acoustic waves within the Pt film, which propagates into BP. The acoustic wave front (dashed planes) serves as a grating moving downwards (blue arrow) that reflects the penetrated probe beam. The reflected probe A (at the Pt/air interface) and B (by the moving grating) will interfere optically, resulting in oscillations in the Brillouin Scattering signals.

Figure 1b shows a front-pump front-probe ps-interferometry schematic for measurements of a BP flake as the representative sample substrate. The probe laser partially penetrates through the optically semitransparent Pt transducer and partially gets reflected at the Pt/air interface (reflected probe A). The acoustic pulse, moving at the speed of sound in the BP sample, changes its local optical constants (*n* and *κ*) through the nonlinear acousto-optic effect.[48] This local change in the optical properties of BP acts as a moving grating (dashed planes in Figure 1b) which can scatter light, and is also commonly known as Brillouin scattering.[49] When the transmitted probe laser arrives at the instantaneous location of the acoustic wave, it will be partially reflected back due to BS, as the reflected probe B shown in Figure 1b. These two reflected probe beams (A and B) can interfere with each other constructively or destructively depending on the instantaneous location of the acoustic wave. The oscillation of the reflectance signal due to this interference can be captured by the ps-interferometry system (in-phase signal in time-domain thermoreflectance measurements).[46,50-52] To obtain the optical constants of the material via analyzing the interference data, several periods of the oscillating signals need to be collected.

Figure 2 depicts a typical BS signal (black line), performed on the sample stack of 5-nm Pt on BP with the incident light being polarized along the ZZ crystalline direction. The BS signal is plotted after the thermal background ($V_{BG}$, red line in the inset of Figure 2) is subtracted from the original signal ($V_{in}$, black line in the inset of Figure 2). It should be noted that we use a rapid scan with a time constant of 0.7 s over a long-range of time delay (up to 2 ns) for data collection to generate Figure 2. For ps-interferometry measurements to determine the optical constants of BP (as shown in Figures 5 and 6 to be discussed later), we use a higher time constant of 1.2 s to improve the signal-to-noise ratio for data collection of up to 300 ps.

As the acoustic wave propagates through the entire BP flake and reaches the interface between BP flake and thermal tape, the acoustic wave is reflected at the interface with an acoustic wave reflection coefficient $r_{ac} = (Z_s - Z_t)/(Z_s + Z_t)$, where $Z_t$ and $Z_s$ are the acoustic impedance (defined as the product of material density and speed of sound) of the thermal tape and BP sample, respectively. The lower acoustic impedance of the thermal tape will cause a $\pi$ shift in the acoustic wave at the interface, and the phase shift will produce a negative peak in the thermal background signal. As shown in the inset of Figure 2, the negative peak at ~1340 ps indicates that the acoustic wave has returned to the top surface. This can also be confirmed by the bumpy feature of the BS signal at ~1340 ps in Figure 2, which results from the enhanced interference signal when the acoustic wave front moves back to the top surface.

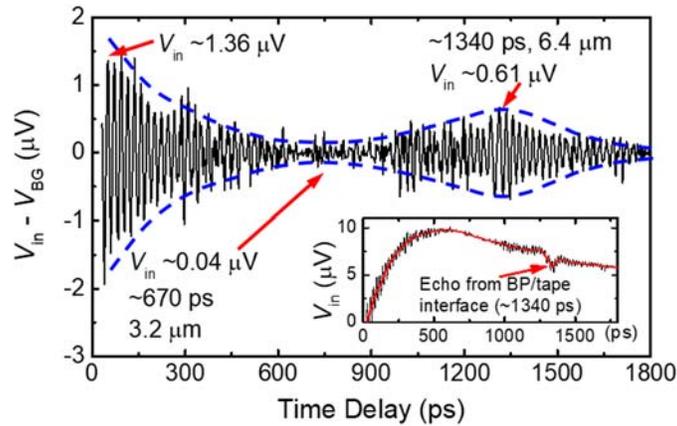

**Figure 2.** A representative BS signal from 5-nm Pt/BP after subtracting the thermal background ($V_{BG}$, red line in the inset) from the original ps-interferometry signal ($V_{in}$, black line in the inset). The incident light is polarized along the ZZ direction. The blue dashed lines indicate the boundaries of the BS wave packet. The amplitude of BS oscillations reaches the minimum at ~670 ps, due to the strongest BP absorption of Probe B before it gets reflected by the acoustic wave traveling to the BP/tape interface. When the acoustic wave returns to the top surface at ~1340 ps, the amplitude of BS oscillations increases again due to the reduced optical path of Probe B in BP. The negative peak of the red curve at ~1340 ps in the inset of Figure 2 indicates that the acoustic wave has reached the BP/tape interface and is reflected to the sample surface during one measurement.

## 2.2. Theoretical Model

The physics of ps interferometry has been well-understood for non-absorbing media.[43,45,47] The frequency and attenuation rate of the oscillating signal, shown in Figure 2, depends on the optical properties of the sample. For a normal incident angle, the change in reflection $\Delta R$ has two components that can be described as a function of time delay and the material properties by

$$\Delta R = |r_\text{o} + r_\text{sound}|^2 = |r_\text{o}|^2 + 2|r_\text{o}||r_\text{sound}|\cos(\frac{2\pi n v_\text{s} t}{\lambda} - \phi) + |r_\text{sound}|^2, \tag{1}$$

where $r_\text{o}$ and $r_\text{sound}$ are the Fresnel reflection coefficients of the probe metal transducer reflection and the probe acoustic wave reflection, $t$ is time, $v_\text{s}$ is the speed of sound, $\phi$ is the phase shift, and $n$ is the refractive index. The term $|r_\text{o}|^2$ is the change in thermoreflectance signal caused by the excitation process, and the middle term on the right-hand side of Eq. 1 is the oscillation in thermoreflectance signal caused by the constructive or destructive interference in the reflected probe laser. The oscillation frequency is directly related to the speed of sound in the substrate medium and the refractive index. In order to separate the two reflection components and focus solely on analyzing the oscillation, the excitation process signal is extracted by applying a second-order Savitzky-Golay smoothing filter with points in a window of delay-time longer than the oscillation period. The thermal background signal due to pump excitation is then subtracted from the original signal to obtain the pure BS oscillation signal. For bulk absorbing media such as BP, light absorption contributes to the attenuation of the oscillation amplitude. In this case, the full-level of oscillation attenuation includes both acoustic wave decay and light absorption, which can be expressed as

$$\Delta R = S(\lambda)\exp(-\beta_\text{D} t)\exp(-\beta t)\cos(2\pi n v_\text{s} t/\lambda - \phi), \tag{2}$$

where $S(\lambda)$ is the BS coefficient, $\beta_\text{D}$ is the acoustic wave decay rate, and $\beta$ is the optical absorption decay rate, which relates with the absorption coefficient $\alpha$ and the speed of sound as $\beta = 2\alpha v_\text{s}$.[43] By fitting the oscillation signals from ps interferometry, the extinction coefficient $\kappa$ can be obtained via $\alpha = 4\pi\kappa/\lambda$.

The two exponential decay constants ($\beta$ and $\beta_\text{D}$) cannot be separately observed in the experiment even though they are caused by different mechanisms. However, the exponential decay of the acoustic wave in the propagation medium is usually very slow in comparison to the time scale of the measurement. The slowly decaying acoustic wave was observed in the BS oscillation signals of glass and SiO$_2$ in previous studies,[42-43] suggesting the signal decay due to the acoustic wave damping can be neglected. In our experiments, we have also verified this assumption in BP. When the probe light polarization is aligned along the ZZ direction, BP should exhibit the least amount of light absorption. Thus, the impact of acoustic wave decay should be more significant in the attenuation in this case. However, as shown in Figure 2, compared to the amplitude of the first oscillation (1.36 $\mu$V), the 45% of the oscillation remains at the time delay of ~ 1340 ps (0.61 $\mu$V). This 55% reduction in the oscillation amplitude at ~1340 ps includes the contributions from both the mechanical damping of the acoustic wave and the acoustic wave transmission at the BP/tape interface. The mechanical damping decay is an order of magnitude lower than the light absorption decay, where only 3% of the oscillation is captured at ~670 ps (0.04 $\mu$V). The optical penetration depth is estimated to be 3.2 $\mu$m, corresponding to a time delay of 670 ps. This confirms that the acoustic wave exponential decay rate is much slower than the optical absorption decay rate in BP. Therefore, we omit the acoustic wave decay rate $\beta_D$ in our data analysis and only consider the decay due to optical absorption.

## 2.3. Sensitivity Analysis

To better illustrate the effectiveness of the optical property measurement through ps-interferometry, we calculate the sensitivity of the measurement signals to different parameters. In ps interferometry, since the measurement signal is oscillating in nature, the sensitivity is defined as the normalized signal $S(x) = \partial\left[-V_{in} / \max(V_{in})\right]/(\partial x / x)$, where $V_{in}$ is the in-phase signal and $\max(V_{in})$ is the maximum value of the measured $V_{in}$, taken as the amplitude of the first BS oscillation peak. $x$ represents one of the parameters that the measurement is sensitive to, including the optical decay $\beta$, oscillation frequency $f$, and the phase shift $\phi$. In the calculations of $S$, a small perturbation of 0.2% is introduced for all the $\partial x / x$, and the calculated $S$ does not change when this perturbation is further reduced.

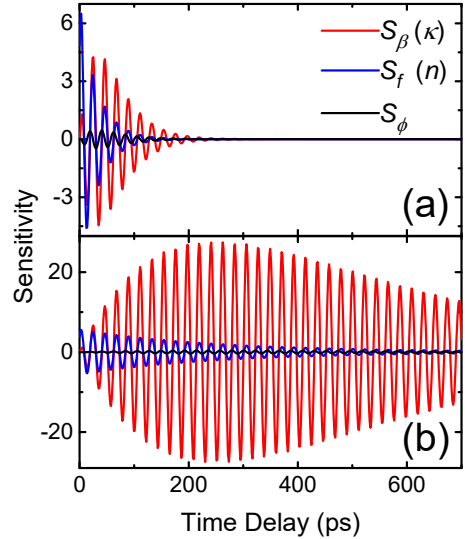

**Figure 3**. The ps-interferometry signal sensitivities to optical absorption decay $\beta$ (or $\kappa$, red line), frequency $f$ (or $n$, blue line), and phase shift $\phi$ (black line). The value of $\kappa$ used in the plots are (a) 0.21 and (b) 0.04, respectively, as two representative values of the extinction coefficient of BP along the AC and ZZ directions.

Examples of the calculated sensitivities are plotted as functions of time delay in Figure 3. The value of $n$ is set as 3.8 (from the averaged frequency of the oscillating BS signals) for all the sensitivity analysis, while the values of $\kappa$ are 0.21 (AC) and 0.04 (ZZ) in Figures 3a and b, respectively. It can be clearly seen that the sensitivity itself also oscillates, and the measurement is very sensitive to both the frequency $f$ (related to the refractive index $n$) and the phase shift $\phi$. The signal of ps interferometry is less sensitive to the extinction coefficient $\kappa$ compared to $f$ (or $n$) and $\phi$. When $\kappa$ is lower, the sensitivity to $\beta$ (or $\kappa$) maintains a certain value throughout the first 1000 ps of time delay as shown in Figure 3b (red line). This is favorable to the extraction of a precise $\kappa$ value. For the higher-$\kappa$ case in Figure 3a, though the sensitivity to $\kappa$ (red line) attenuates to almost zero before 200 ps of time delay, the amplitudes of these sensitivity oscillations at the first few periods are apparently higher than those of the lower-$\kappa$ case, which already provides sufficient data for a reliable fitting. Thus, the sensitivity analysis justifies the robustness of our ps-interferometry method for determining the optical properties regardless the high or low values of $\kappa$ for various materials.

The duration of time delay at which the measurements have sufficient sensitivities also determines the necessary sample thickness to gather enough data via ps interferometry. For example, in both cases plotted in Figure 3, the periods of the wavy feature and the signal sensitivity to $\beta$ (same to $\kappa$) are both sufficiently high during the first 200 ps. This suggests that a sample thickness of ~1 $\mu$m, obtained as the product of time duration (~200 ps) and speed of sound (~5 nm ps$^{-1}$), is adequate to gather useful data for reliable measurements.

Compared with conventional spectroscopic approaches,[53-55] ps interferometry has several advantages. The thin transducer for launching acoustic waves naturally serves as a protective layer, preventing BP from rapid degradation. It can directly measure the optical constants of anisotropic materials and extract both $n$ and $\kappa$ simultaneously from reflectance only. Thus, optical absorption measurements by ps interferometry do not require information about the sample thickness, as typically it can directly probe $\kappa$ without knowing the optical transmission of the sample.[45]

## 3. Results and Discussion

### 3.1. Sample Structure

Single crystals of bulk BP are synthesized using a low-pressure synthesis technique.[56] In Figure 4a, no impurity phase is observed in the X-ray diffraction (XRD) patterns, which suggests the presence of only a single phase of BP. Only the (0 k 0) reflections can be recognized (k = 2, 4, or 6), which confirms that the crystal surface normal direction is parallel to the b axis, and the sheets are parallel to the a–c plane. To determine the in-plane orientations of BP, we measured the [061] planes using XRD following the approach detailed in Ref. [20]. The results are consistent with literature studies showing that the narrow width of the ribbon-like BP flake is along the armchair direction (c axis), and the longer length of the ribbon-like BP flake corresponds to the zigzag direction (a axis). Figure 4b illustrates an (Atomic Force Microscope) AFM image scanned over a 3 μm × 3 μm area with a root-mean-square (rms) roughness of 0.9 nm. Figure 4c shows the BP ribbon exfoliated from the bulk BP (inset picture) with notation of the in-plane crystalline directions. Figure 4d depicts an optical image of the measured BP sample surface, which suggests a long-range flatness.

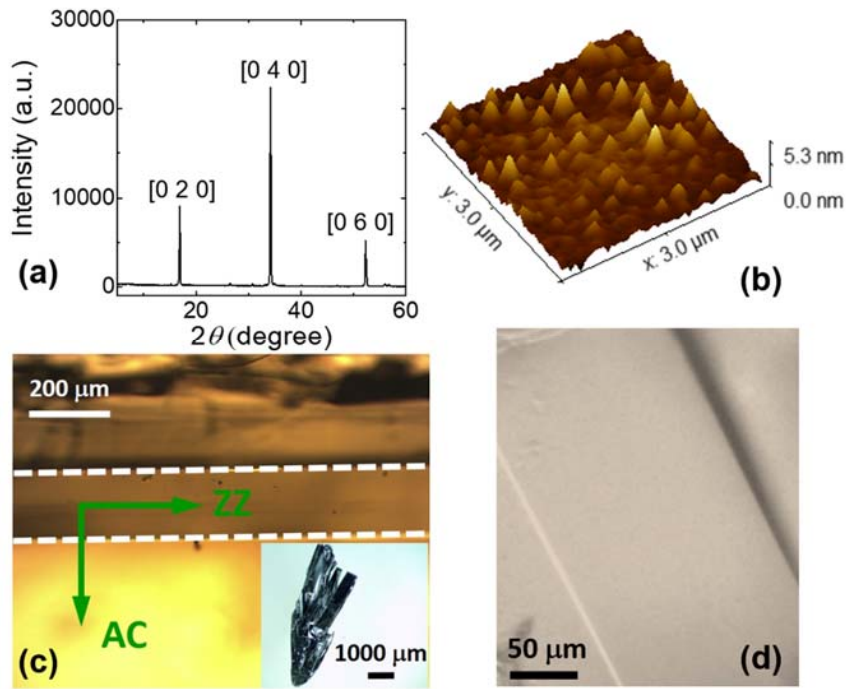

**Figure 4.** (a) The powder XRD $2\theta$ scan of the BP sample revealing the [0 2 0], [0 4 0] and [0 6 0] crystalline planes. (b) The AFM image of BP flake on a Si substrate. (c) Optical microscopic image of the ribbon-like BP flake. The dashed lines indicate the BP flake exfoliated from the bulk crystal

(inset). (d) The microscopic image of the measured BP sample surface, which suggests a long-range flatness.

### 3.2. Polarization-Dependent Optical Constants of BP

BP has been observed to have the highest (lowest) optical absorption when the incident light is linearly polarized aligned along its AC (ZZ) direction.[3,19,27-28] In this experiment, the light polarization along the ZZ direction is defined as 0°, and that along the AC direction is defined as 90°. When the linearly polarized light is aligned along either the ZZ or AC crystal direction, the optical constants can be directly extracted from the experimental data based on Eq. 2. Using the reported speed of sound value 4.76 nm ps$^{-1}$ in BP along the interlayer direction,[57-58] the extracted refractive index $n$ is $3.77 \pm 0.19$ at a wavelength of $\lambda = 783$ nm, which agrees well with reported values of 3.83 from literature (within 2% difference).[29] And the values of $n$ show very weak dependence on incident light polarization, which confirms the previous observations on the nanometer-thick BP films with a light wavelength in the near infrared.[3,26] At the same time, the absorption coefficients $\alpha$ along the ZZ and AC directions can also be extracted by fitting the BS signals with Eq. 2. A strong anisotropy of BP absorption is observed as consistent with previous theoretical predictions and indirect measurements of BP optical constants.[3,39] We further determine the extinction coefficients $\kappa$ for both directions using $\alpha$, and the fitted $\kappa_{ZZ}$ and $\kappa_{AC}$ are $0.027 \pm 0.008$ and $0.210 \pm 0.063$ respectively. These results on micron-thick BP are consistent with the reported values on 100~200-nm film of 0.017 and 0.210 at this wavelength that were indirectly obtained via EELS measurements.[29]

At an arbitrary polarization angle between 0° and 90°, Eq. 2 is not sufficient to describe the birefringent behavior of BP, and the combined absorption of $\kappa_{ZZ}$ and $\kappa_{AC}$ must be considered. For linearly polarized light oriented at an arbitrary angle $\theta$, the intensity of the reflected wave is $S = |E|^2/2\eta_0$ where $\eta_0$ is the free space impedance and $E$ is the electric field as a function of the incident polarization angle. The reflected electric field can be decomposed into two polarization components as $E_r = e_x E_x + e_y E_y = |E_r|(e_x \cos\theta + e_y \sin\theta)$. The sum of $x$ and $y$ direction intensity components is $S = (|E_{rx}|^2 + |E_{ry}|^2)/2\eta_0$. In this case, the change in reflectivity at an arbitrary polarization angle $\theta_p$ is

$$\Delta R(\theta_p) = \Delta R(0)\cos^2\theta + \Delta R(90)\sin^2\theta. \tag{3}$$

Experimental data can be fitted by substituting the polarization angle, $\kappa_{ZZ}$ and $\kappa_{AC}$ into Eq. 3, which becomes a bi-exponential function similar to Eq. 2:

$$\begin{aligned}\Delta R(\theta_p) = S(\lambda)[&\cos^2\theta_p \exp(-2\alpha_{ZZ}v_s t)\cos(2\pi n v_s t/\lambda - \phi) \\ + &\sin^2\theta_p \exp(-2\alpha_{AC}v_s t)\cos(2\pi n v_s t/\lambda - \phi)]\end{aligned} \tag{4}$$

Figure 5a shows the BS signals taken at various incident polarization angles between 0° and 90°, with a higher time constant of 1.2 s to improve the signal-to-noise ratio for data collection. The solid lines for 0° and 90° are fitted with Eq. 2 excluding the acoustic wave decay term. The solid lines for any random angles other than 0° and 90° are the fitted with Eq. 4. The amplitude of BS oscillations gradually decreases as the measurement angle changes from 0° to 90°. This indicates that more light is absorbed when polarization direction is aligned along the AC direction and less interference can be observed. $A(\theta_p, t)$, the change in the oscillation amplitude of $\Delta R$ can be described as

$$A(\theta_p, t) = \cos^2\theta_p \exp(-2\alpha_{ZZ}v_s t) + \sin^2\theta_p \exp(-2\alpha_{AC}v_s t), \tag{5}$$

where $\alpha_{ZZ}$ and $\alpha_{AC}$ are $6.42\times10^{-4}$ nm$^{-1}$ and $3.37\times10^{-3}$ nm$^{-1}$, respectively. The oscillation decay of the original signals is compared with the bi-exponential decay of Eq. 5 in Figure 5b. At each incident light polarization angle, the oscillation decay of the signal can be expressed as a function of the absorption coefficients along both the ZZ and AC polarization directions. This is an excellent evidence of BP's linear dichroic characteristics with anisotropic optical absorption, which is a desired property for future polarization-sensitive optoelectronic devices, such as photodetectors, field-effect transistors (FET), LEDs, solar cells, displays or optical switches.

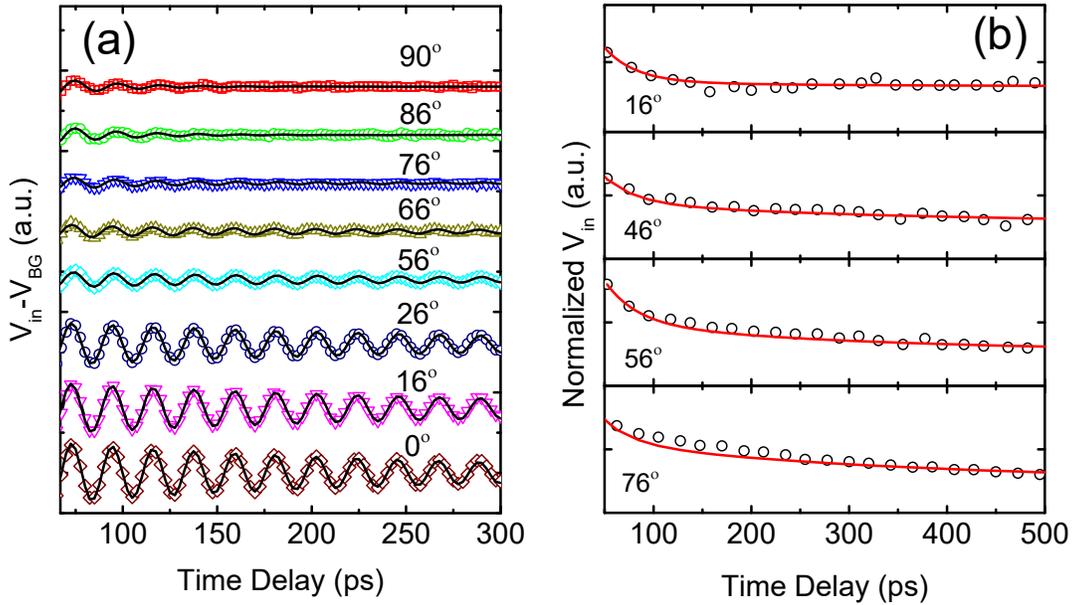

**Figure 5.** Picosecond interferometry signals from BP. (a) The signals subtracted with thermal background $V_{BG}$ at various angles between the ZZ and AC crystalline directions (open symbols). The solid lines represent the best fittings with the models. (b) The amplitude of the oscillations as a function of time delay. The signals are normalized to the first peak. The black open circles are the amplitude of oscillation of the original data, and the red solid lines are bi-exponential calculations from Eq. 5 containing the angular-dependent decaying term.

### 3.3. Optical Constants of BP Measured at Longer Wavelengths

The ps-interferometry measurements on BP at 835 nm and 890 nm are also conducted as shown in Figure 6. This is done by tuning the laser wavelength and keeping the wavelength full-width half maximum to be less than 13 nm while maintaining all other measurement parameters at their same values. The $n$ and $\kappa$ values at 835 nm are 3.70 ± 0.19, 0.156 ± 0.047 for the AC direction, and 3.73 ± 0.19, 0.035 ± 0.011 for the ZZ direction. At 890 nm, *n* and *κ* are 3.69 ± 0.19, 0.267 ± 0.080 for the AC direction, and 3.69 ± 0.19, 0.032 ± 0.010 for the ZZ direction. The results demonstrate BP's strong linear birefringence of optical absorption in the near infrared region.

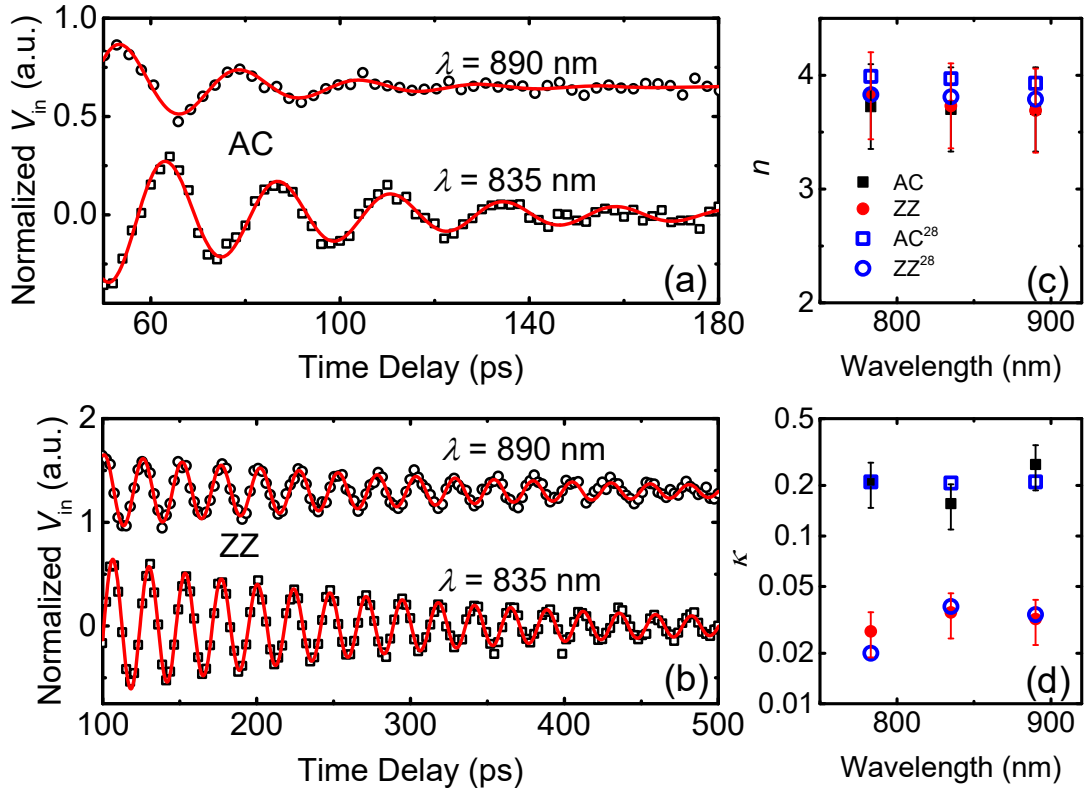

**Figure 6.** The oscillation signals (discrete points) along the AC (a) and ZZ (b) directions obtained at 835 nm and 890 nm. By fitting the experiment data with Eq. 2 (red solid lines), *n* (c) and *κ* (d) values are extracted. Black solid squares are the measured values in the AC direction and solid red circles are the measured values in the ZZ direction. Open blue squares and circles are literature values[29] of the optical constants for BP thin films along AC and ZZ, respectively. The error bars for obtained *n* and *k* in this work are 10% and 30%, respectively.

As shown in Figures 6c and 6d, our ps-interferometry data agree reasonably well with the previous EELS results,[29] with a difference within our measurement uncertainties (note that measurement error bars are not provided in Ref. [28]). This deviation is presumably caused by the difference between these two measurement methods, and the sample system variations. In addition, the thicknesses of BP flakes in previous EELS measurements ranged from 100 to 200 nm,[29] while the thicknesses of our samples are on the order of a few microns. Since BP is known for its thickness-dependent optical properties, this difference in sample thickness may also account for discrepancies in the reported values for the optical constants of BP.[38] Here, we also would like to point it out that we only present the discrete measurement data within near-IR spectrum due to the limit of operating wavelength range of our Ti-Sapphire ultrafast laser (from 760 to 900 nm). However, the approach of ps interferometry for studying materials' optical properties can be extended to wider spectra by incorporating a laser source with a broad tunable wavelength range[59] into the system.

## 4. Conclusion

In summary, a ps-interferometry method is developed to probe the polarization-dependent optical properties of micrometer-thick BP. This method naturally provides BP with metal-layer protection against the sample's degradation caused by the ambient environment, allowing for the direct measurements of optical properties for the materials of interest with high accuracy. The ps-interferometry signals of BP along the through-plane direction with different probing wavelengths (780-890 nm) and polarization directions are obtained. The optical extinction coefficient with the probe beam being polarized along the AC direction of BP (~0.2) is much higher than that along the ZZ direction (~0.03). This confirms the linear birefringence of BP at the wavelengths used in this study. Meanwhile, the refractive index $n$ remains almost constant throughout the range of the 780 to 890 nm wavelengths. For linearly polarized incident light that is not aligned along either the AC or ZZ crystalline direction, the ps-interferometry signals can be well described by a newly-developed bi-exponential decay model. Both the measurement results and sensitivity analysis show that the ps-interferometry method is robust for optical measurements on materials with a wide range of optical properties and thicknesses. The optical absorption data and approach presented in this work would benefit engineering of future applications of BP. A laser with a broad tunable spectrum would be able to easily generalize our ps-interferometry approach for studying the optical properties of other 2D materials over a wide range of optical wavelengths.

## 5. Experimental Section

*Sample preparation*: High-quality single crystals of BP are grown with a low-pressure synthesis technique, which has been described previously.[20] A mixture of 60 mg of Sn (Alfa Aesar, 99.85%), 30.6 mg of $SnI_4$ (Alfa Aesar, 99.998%), and 1.5g of red phosphorus (Alfa Aesar, 98.9%) are vacuum sealed in a quartz tube with an inner diameter of 12 mm and a length of 25 cm. The sealed quartz ampoule is placed inside a tube furnace and heated to 650 °C. After maintaining the temperature at 650 °C for 1 hour, the furnace is cooled for 26.2 hours at a constant rate of 0.22 °C per minute down to 300 °C, during which time the BP crystals are formed. The BP crystal structure is examined with powder X-ray diffraction from a Rigaku Smartlab 9 KW with Cu K-α radiation ($\lambda$=1.5406 Å). The BP sample grown by this technique tends to be exfoliated along the ZZ direction. The bulk sample is mechanically exfoliated into smaller flakes and fixed onto a silicon substrate with a thermally conductive tape.

*Optical measurement method*: The picosecond interferometry setup used in this study is modified from the basic pump-probe setup, which has been described elsewhere.[60-61] The pump and probe laser beams have a pulse duration of ~0.1 ps and the wavelength are centered at 783, 835, and 890 nm in this study. The laser repetition rate is 80 MHz. The pump laser is modulated by an electro-optic modulator at 9 MHz, and the probe laser is modulated by a mechanical chopper at 200 Hz. Since the sample flake is small, a 10× objective lens is used to focus the pump laser and the probe laser onto the sample, which produces $1/e^2$ beam spot radius of approximately 6 μm. The pump laser power and probe laser power are 5 mW and 5 mW, respectively. To measure the anisotropic optical properties of BP, a half waveplate mounted on a high precision rotation mount is placed before the focusing lens to tune the polarization direction of the probe laser. When the waveplate is rotated by an arbitrary angle ($\theta$), the probe laser polarization angle is rotated by $\theta_p = 2\theta$. The 1° resolution of the optical mount produces a rotational resolution of 2° for actual measurements. The reflected probe signal is received by a photodetector and sent to a lock-in amplifier for post data processing.